\documentclass{article}
\usepackage{amsmath}
\usepackage{amscd}
\usepackage{graphicx}
\usepackage{latexsym}
\usepackage{amsmath,amsthm,amsfonts,amssymb,cite%
}
\usepackage{latexsym}
\usepackage{amsmath}

\theoremstyle{remark}

\newtheorem*{rem}{Remark}
\renewcommand\dagger*
\newcommand\Rho{P}
\renewcommand\Theta{\boldsymbol\theta}
\numberwithin{equation}{section}
\begin{document}
\hoffset = -2.4truecm \voffset = -2truecm
\renewcommand{\baselinestretch}{1.2}
\newcommand{\mb}{\makebox[10cm]{}\\ }
\title{On solutions to the non-Abelian Hirota-Miwa equation and its continuum limits}
\author{C.X. Li$^{1}$, J.J.C. Nimmo$^{2}$ and K.M. Tamizhmani$^{3}$
\\
$^{1}$School of Mathematical Sciences,\\
Capital Normal University\\
Beijing 100048, CHINA\\
$^{2}$Department of Mathematics, \\
University of Glasgow \\
Glasgow G12 8QW, UK\\
$^{3}$Department of Mathematics,\\
Pondicherry University\\
Kalapet, 605014 Pondicherry, India }
\date{}
\maketitle

\begin{abstract}
In this paper, we construct grammian-like quasideterminant solutions of a
non-Abelian Hirota-Miwa equation. Through continuum limits of
this non-Abelian Hirota-Miwa equation and its quasideterminant
solutions, we construct a cascade of noncommutative differential-difference equations ending with the noncommutative KP equation. For each of these systems the quasideterminant solutions are constructed as well.
\end{abstract}

\section{Introduction}
Recently, noncommutative versions of some well-known soliton
equations have been extensively studied. These systems include the
KP equation, the KdV equation, the Hirota-Miwa equation, the
modified KP equation and the two-dimensional Toda lattice equation
\cite{K,P,S,WW1,WW2,WW3,H,HT1,DH,JN,GN1,GN2,GNS,LN}. Generally
speaking, determinants are ubiquitous as solutions of commutative
integrable systems while quasideterminants are often solutions of
noncommutative integrable systems \cite{GR,EGR,GGRL}. In many cases,
the noncommutative version of the equation is obtained as the
compatibility of the same Lax pair for the commutative version.
However, in the noncommutative case, the assumption that the
coefficients in the Lax pair commute is relaxed.

This paper is concerned with the non-Abelian version of the Hirota-Miwa equation
introduced in \cite{JN}. Given the system of linear equations
\begin{equation}
a_i^{-1}T_i(\phi)-a_j^{-1}T_j(\phi)=U_{ij}\phi,\quad
i,j\in\{1,2,3\},\label{LPHM}
\end{equation}
where $\phi$ and $U_{ij}$ belong to an associative algebra $\mathcal
A$ and depend on variables $n_1,n_2,n_3$, $T_i$ denotes the shift
operator in variable $n_i$, defined by $T_i(X)=X(n_i+1)$ and $a_i$
are nonzero scalar constants, called lattice parameters. We will
also use the shorthand subscript notation $X_{,i}=T_i(X)$. This
linear system is compatible if and only if
\begin{align}
&U_{ij}+U_{jk}+U_{ki}=0,\label{HM1}\\
&T_k(U_{ij})+T_i(U_{jk})+T_j(U_{ki})=0,\label{HM2}\\
&T_k(U_{ij})U_{ki}=T_j(U_{ki})U_{ij},\label{HM3}
\end{align}
for each $i,j,k\in\{1,2,3\}$. In particular, these give $U_{ii}=0$ and $U_{ij}=-U_{ji}$ for $i,j\in\{1,2,3\}$. Equations \eqref{HM1}-\eqref{HM3} are referred to as the non-Abelian Hirota-Miwa equation \cite{JN,GN2} since in the commutative case if we make the ansatz
\begin{equation}\label{U}
    U_{ij}=(a_i^{-1}-a_j^{-1})\dfrac{T_{ij}(\tau)\tau}{T_i(\tau)T_j(\tau)},
\end{equation}
where $T_{ij}(\tau)$ denotes $T_i(T_j(\tau))$, then all of the
equations in \eqref{HM1}--\eqref{HM3} reduce to just one equation,
the well-known Hirota-Miwa equation \cite{HR,MT}
\begin{equation}\label{HM}
 a_1(a_2-a_3)T_1(\tau)T_{23}(\tau)+a_2(a_3-a_1)T_2(\tau)T_{31}(\tau)+a_3(a_1-a_2)T_3(\tau)T_{12}(\tau)=0.
\end{equation}

In the commutative case, it is well known that the Miwa
transformation \cite{MT}
\[
x_k=\sum_{i=1}^{\infty}\frac{a_i^k}{k}n_i,
\]
provides the link between the discrete variables $n_i$ and the continuous variables $x_k$ in the KP hierarchy. Using this transformation, we can view $\tau=\tau(n_1,n_2,n_3,\dots;x_1,x_2,x_3,\dots)$ as a function of both discrete and continuous variables such that
\begin{equation}
    T_i(\tau)=\left.\tau\right|_{n_i\to n_i+1}=
    \left.\tau\right|_{x_1\to x_1+a_i,x_2\to x_2+a_i^2/2,x_3\to
    x_3+a_i^3/3,\dots},
\end{equation}
and the continuum limits, as lattice parameters $a_i\to0$, are obtained by using Taylor's theorem. In this way we have, for example,
\[
T_i(\tau)=\tau+a_i\tau_{x}+\tfrac12a_i^2(\tau_{xx}+\tau_{y})+\tfrac16a_i^3(\tau_{xxx}+3\tau_{xy}+2\tau_{t})+O(a_i^4),
\]
where, for a shorter notation, we have written $x_1,x_2,x_3$ as $x,y,t$ respectively.

\begin{figure}[htb]
    \begin{center}
    \setlength{\unitlength}{1mm}
    \begin{picture}(90,11)
    \thicklines
 \put(-28,0){\framebox(23,10)}\put(-5,5){\vector(1,0){18}}
 \put(13,0){\framebox(23,10)}\put(36,5){\vector(1,0){18}}\put(54,0){\framebox(23,10)}\put(77,5){\vector(1,0){18}}
 \put(95,0){\framebox(23,10)}
 \put(-26,4){Hirota-Miwa}\put(19,4){$\Delta^2\partial$KP}\put(60,4){$\Delta\partial^2$KP}\put(104,4){KP}
  \put(-2,6){$a_3\rightarrow 0$}\put(39,6){$a_2\rightarrow 0$}\put(80,6){$a_1\rightarrow 0$}
    \end{picture}
    \end{center}
    \caption{Continuum limits\label{fig1}}
\end{figure}
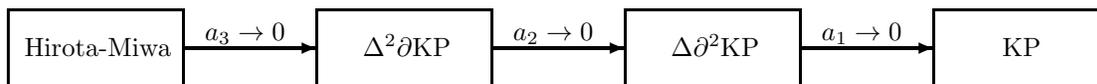

Taking successive continuum limits one obtains a cascade of
differential-difference equations starting from the Hirota-Miwa
equation and ending with the KP equation. To be precise, as
$a_3\to0$ the leading order term in \eqref{HM} is at $O(a_3)$ and is
\begin{equation}\label{Delta^2KP}
    (a_1-a_2)(T_1(\tau)T_2(\tau)-T_{12}(\tau)\tau)+a_1a_2D_xT_1(\tau)\cdot T_2(\tau)=0,
\end{equation}
which we call the $\Delta^2\partial$KP equation. Next, as $a_2\to0$ the leading order term in \eqref{Delta^2KP} is at $O(a_2^2)$ and is
\begin{equation}\label{DeltaKP}
    (a_1D_y+a_1D_x^2-2D_x)T_1(\tau)\cdot\tau=0,
\end{equation}
which is nothing but the differential-difference KP equation. Here
we denote \eqref{DeltaKP} as the $\Delta\partial^2$KP equation to
stress that it has one discrete variable and two continuous ones.
Finally, as $a_1\to0$ the leading order term in \eqref{Delta^2KP} is
at $O(a_1^3)$ and is
\begin{equation}
    (D_x^4+3D_y^2-4D_xD_t)\tau\cdot\tau=0,
\end{equation}
the Hirota form of the KP equation. The whole procedure is
illustrated in Figure~\ref{fig1}.

The main aim of this paper is to show that the same cascade of continuum limits also exists in the non-Abelian case and thereby we construct noncommutative versions of the differential-difference KP equations \eqref{Delta^2KP} and \eqref{DeltaKP}. We also give solutions of these systems in terms of quasideterminants.

The paper is organized as follows. In Section~2, we review results from \cite{JN} on the Daroux transformations for the non-Abelian Hirota-Miwa equation and consider the binary Darboux transformation and the quasigrammian solutions of \eqref{HM1}--\eqref{HM3} that it constructs. Using the ansatz that is suggested by this binary Darboux transformation, in Section~3, we first reformulate the
results presented in a way which facilitates taking continuum
limits. Then, by taking the successive continuum limits described
above we first obtain a noncommutative version of \eqref{Delta^2KP}.
Quasiwronskian solutions and quasigrammian solutions for this system
are also obtained. Then a second continuum limit yields a
noncommutative version of \eqref{DeltaKP} and its quasiwronskian and
quasigrammian solutions are presented. Finally we show that the
continuum limit of this noncommutative differential-difference KP is
nothing but the noncommutative KP equation. Conclusions and
discussions are given in Section~4.

\section{The non-Abelian Hirota-Miwa equation}
In this section, we will present quasigrammian solutions to the
non-abelian Hirota-Miwa equation \eqref{HM1}--\eqref{HM3}
constructed by using binary Darboux transformations.

Let us first review some results on the non-Abelian Hirota-Miwa
equation. Quasicasoratian solutions were constructed for the
non-Abelian Hirota-Miwa equation by repeating Darboux
transformations in \cite{JN,GN2}. The linear system \eqref{LPHM}
considered here is a generalisation of the one used in \cite{JN,GN2}
but it is not difficult to prove that \eqref{HM1}--\eqref{HM3} has
the Darboux transformation
\begin{align}
\tilde \phi&=a_i^{-1}(T_i(\phi)-T_i(\theta)\theta^{-1}\phi),\label{HMDT1}\\
\tilde U_{ij}&=T_j(T_i(\theta)\theta^{-1})U_{ij}\theta T_i(\theta)^{-1},\quad
i,j\in\{1,2,3\}\label{HMDT2}
\end{align}
where $\theta$ is a particular solutions of the linear system
\eqref{LPHM}. This suggests the ansatz $U_{ij}=T_j(C_i)U^0_{i,j}C_i^{-1}$, where $U^0_{i,j}$ is a seed solution. Then iterating the Darboux transformation gives casoratian-type quasideterminant solutions
\begin{equation}\label{DT C}
    C_i[n]=\begin{vmatrix}
        \Theta&1\\
        T_i(\Theta)&0\\
        \vdots&\vdots\\
        T_i^{n-1}(\Theta)&0\\
        T_i^n(\Theta)&\fbox{$0$}
    \end{vmatrix}C_i,
\end{equation}
where $\Theta=(\theta_1,\dots,\theta_N)$ is a vector solution of \eqref{LPHM}. In the next section we will rewrite this Darboux transformation using a different parametrization and obtain a different casoratian-type solution.

To construct binary Darboux transformations for
\eqref{HM1}--\eqref{HM3}, we must define an appropriate adjoint
system. The formal adjoint of the linear system \eqref{LPHM} has the
form
\[
a_i^{-1}T_i^{-1}(\psi)-a_j^{-1}T_j^{-1}(\psi)=V_{ij}\psi,\quad
i,j\in\{1,2,3\}.
\]
and we will choose $V_{ij}$ in order that the compatibility condition of this
adjoint system is equivalent to \eqref{HM1}--\eqref{HM3}. The compatibility condition for this system of linear equations can be deduced from \eqref{HM1}--\eqref{HM3} by simply
changing $T_i$ to $T_i^{-1}$ and $U_{ij}$ to $V_{ij}$ in this system and
we get
\begin{align}
&V_{ij}+V_{jk}+V_{ki}=0,\label{aHM1}\\
&T_k^{-1}(V_{ij})+T_i^{-1}(V_{jk})+T_j^{-1}(V_{ki})=0,\label{aHM2}\\
&T_k^{-1}(V_{ij})V_{ki}=T_j^{-1}(V_{ki})V_{ij}.\label{aHM3}
\end{align}
This is equivalent to \eqref{HM1}--\eqref{HM3} when $T_{ij}(V_{ij})=U_{ij}^*$, where $^*$ is an involutive antiautomorphism defined on $\mathcal A$ satisfying $(AB)^*=B^* A^*$ and $(A^*)^*=A$. Thus we obtain the adjoint linear system in its final form
\begin{equation}
a_i^{-1}T_j(\psi)-a_j^{-1}T_i(\psi)=U^*_{ij}T_{ij}(\psi).\label{AFHM}
\end{equation}

An eigenfunction potential $\Omega(\phi,\psi)$ may be defined by
\begin{equation}
\Delta_i(\Omega)=T_i(\psi^*)\phi,\label{HMQ}
\end{equation}
for all $i$, where $\Delta_i=a_i^{-1}(T_i-1)$. It is straightforward to show that the exactness condition $(\Omega_{,i})_{,j}=(\Omega_{,j})_{,i}$ when $\phi$ and $\psi$ satisfy \eqref{LPHM} and \eqref{AFHM} respectively.
Following the standard construction, a binary Darboux transformation is then defined by
\begin{align}
\label{HMBDT1}
\tilde{\phi}&=\phi-\theta\Omega(\theta,\rho)^{-1}\Omega(\phi,\rho),\\
\tilde{\psi}&=\psi-\rho(\Omega(\theta,\rho)^\dagger)^{-1}\Omega(\theta,\psi)^*,\label{HMBDT2}\\
\label{HMBDT3}
\widetilde{U}_{ij}&=U_{ij}-(T_i-T_j)(\theta\Omega(\theta,\rho)^{-1}\rho^*),
\end{align}
where $\theta$ and $\rho$ are particular solutions of the linear system \eqref{LPHM} and its adjoint \eqref{AFHM}.

There is also a `vector' form of this transformation, equivalent to an iterated transformation which gives solutions
\begin{equation}
\label{U BDT}
U_{ij}=U^0_{ij}-(T_i-T_j)(\Theta\Omega(\Theta,\Rho)^{-1}\Rho^*),
\end{equation}
where $U^0_{ij}$ is a seed solution of \eqref{HM1}--\eqref{HM3},
$\Theta$ is as defined above and $\Rho=(\rho_1,\dots,\rho_N)$ is a
vector solution of the adjoint linear problem \eqref{AFHM} with
potential $U_{i,j}=U_{ij}^0$. This solution may also be expressed as
a quasideterminant
\begin{align}
U_{ij}=U^0_{ij}+(T_i-T_j)\begin{vmatrix} \Omega(\Theta, \Rho)&\Rho^*\\
\Theta&\fbox{$0$}
\end{vmatrix}.\label{QGHM}
\end{align}

This result can be proved by induction but here we choose to verify the solution by direct substitution into \eqref{HM1}--\eqref{HM3}. We write
\begin{equation}\label{ansatz}
    U_{ij}=U^0_{ij}+G_{,i}-G_{,j},
\end{equation}
where \begin{align} G=\begin{vmatrix} \Omega(\Theta, \Rho)&\Rho^*\\
\Theta&\fbox{$0$}
\end{vmatrix}=-\Theta\Omega^{-1}\Rho^*.\label{QWGHM}
\end{align}
Observe that \eqref{HM1}, \eqref{HM2} are identically satisfied by
the ansatz \eqref{ansatz}, irrespective of the form of $G$, and all
nontrivial versions of \eqref{HM3}, in which $i$, $j$ and $k$ are
distinct, give the same equation
\begin{equation}\label{Gform}
    G_{,12}(U^0_{12}+G_{,1}-G_{,2})+G_{,23}(U^0_{23}+G_{,2}-G_{,3})+G_{,31}(U^0_{31}+G_{,3}-G_{,1})
    +(U^0_{23}G)_{,1}+(U^0_{31}G)_{,2}+(U^0_{12}G)_{,3}=0.
\end{equation}
Using \eqref{HMQ}, we obtain for any $i,j,k$
\begin{equation*}
G_{,ik}G_{,k}=a_i^{-1}\Theta_{,ik}(\Omega_{,k}^{-1}-\Omega_{,ik}^{-1})P_{,k}^*.
\end{equation*}
Then substituting these expressions into \eqref{Gform}, and making use of the linear equations \eqref{LPHM} and \eqref{AFHM}, it is straightforward to show that \eqref{Gform} is identically satisfied. This completes the verification of the solutions \eqref{QGHM}.

\section{Continuum limits of the non-Abelian Hirota-Miwa equation}
In this section, we will find the proper continuum limits of the
non-Abelian Hirota-Miwa equation as the lattice parameters tend to
zero. As well as constructing the nonlinear equations, the
corresponding linear equations (Lax pairs), Darboux and binary
Darboux transformations and solutions will be obtained also in the
same limit.

The first step is to reformulate the results for the non-Abelian Hirota-Miwa equation in a way that facilitates these continuum limits. In particular, we reexpress many terms using the forward and backward difference operators $\Delta_i:=a_i^{-1}(T_i-1)$ and $\nabla_i:=a_i^{-1}(1-T_i^{-1})$. We make an ansatz of the form \eqref{ansatz} suggested by the binary Darboux transformation, in which $U^0_{ij}$ is any seed solution of \eqref{HM1}--\eqref{HM3}. In fact, in order to successfully balance leading order terms, it turns out to be necessary to choose the seed solution in the form $U^0_{ij}=a_i^{-1}-a_j^{-1}$. Thus we write the solution of \eqref{HM1}--\eqref{HM3} as
\begin{equation}
U_{ij}=a_i^{-1}-a_j^{-1}+T_iG-T_jG.\label{QA}
\end{equation}
Under this assumption, the system \eqref{HM1}--\eqref{HM3} reduces to the single equation
\begin{equation}\label{HM4}
\begin{split}
&a_1a_2\bigl(\Delta_{1}\Delta_{2}G(a_1\Delta_1G-a_2\Delta_2G)-[\Delta_1G,\Delta_2G]\bigr)\\
&\quad+a_2a_3\bigl(\Delta_{2}\Delta_{3}G(a_2\Delta_2G-a_3\Delta_3G)-[\Delta_2G,\Delta_3G]\bigr)\\
&\quad\quad+a_3a_1\bigl(\Delta_{3}\Delta_{1}G(a_3\Delta_3G-a_1\Delta_1G)-[\Delta_3G,\Delta_1G]\bigr)=\\
&\quad\quad\quad\quad(a_1-a_2)\Delta_{1}\Delta_{2}G+(a_2-a_3)\Delta_{2}\Delta_{3}G+(a_3-a_1)\Delta_{3}\Delta_{1}G.
\end{split}
\end{equation}
and its linear system and adjoint linear system may be written as
\begin{equation}\label{DDKPLP}
\Delta_i\phi-\Delta_j\phi=(a_i\Delta_iG-a_j\Delta_jG)\phi,
\end{equation}
and
\begin{equation}
-\nabla_i\psi+\nabla_j\psi=(a_i\nabla_iG^*-a_j\nabla_jG^*)\psi,\label{HMAF}
\end{equation}
respectively.

For the purpose of taking continuum limits, it is more convenient to reformulate the Darboux transformation \eqref{HMDT1}--\eqref{HMDT2} using the parametrization \eqref{ansatz} given by the binary Darboux transformation. Doing this we find that
\begin{align}
\tilde\phi&=\Delta_i\phi-(\Delta_i\theta)\theta^{-1}\phi,\label{DTFL1}\\
\tilde G&=T_iG-(\Delta_i\theta)\theta^{-1}.\label{DTFL2}
\end{align}
Superficially, these expressions seem strange as the LHS depends on index $i$ but the RHS does not. It is however easy to check using the linear problem \eqref{DDKPLP} that the RHS in each case is independent of the choice of $i$. Iterating the Darboux transformation in this form again gives quasicasoratian solutions but different from the ones given in \eqref{DT C},
\begin{equation}\label{DT G}
    G[n]=T^n_iG+
    \begin{vmatrix}
        \Theta&0\\
        \vdots&\vdots\\
        \Delta_i^{n-2}\Theta&0\\
        \Delta_i^{n-1}\Theta&1\\
        \Delta_i^n\Theta&\fbox{$0$}
    \end{vmatrix},
\end{equation}
where $\Theta$ is as in \eqref{DT C}. Again, it may be shown that the RHS of \eqref{DT G} is independent of the choice of $i$.

\subsection{Noncommutative $\Delta^2\partial$KP equation}
To take the first continuum limit, let $a_3\to 0$ and note that, for
example, $\Delta_3G=G_x+O(a_3)$.
\paragraph{Nonlinear equation}
The leading order terms in \eqref{HM4} give
\begin{equation}\label{2d1cKP}
a_1a_2\bigl(\Delta_{1}\Delta_{2}G(a_1\Delta_1G-a_2\Delta_2G)-[\Delta_1G,\Delta_2G]\bigr)
=(a_1-a_2)\Delta_1\Delta_2G+a_2\Delta_2G_x-a_1\Delta_1G_x,
\end{equation}
which we call the noncommutative $\Delta^2\partial$KP equation.

\paragraph{Lax pair} The leading order terms in \eqref{DDKPLP} for give the Lax pair
\begin{align}\label{2d1cKPLP1}
    \Delta_1\phi&=\phi_x+a_1(\Delta_1 G)\phi\\
    \label{2d1cKPLP2}
    \Delta_2\phi&=\phi_x+a_2(\Delta_2 G)\phi.
\end{align}
Similarly, the leading order terms in \eqref{HMAF} give the adjoint
form
\begin{align}
    \nabla_1\psi=\psi_x-a_1(\nabla_1G^*)\psi,\label{AFDD1}\\
    \nabla_2\psi=\psi_x-a_2(\nabla_2G^*)\psi.\label{AFDD2}
\end{align}
The compatibility conditions of both \eqref{2d1cKPLP1}--\eqref{2d1cKPLP2} and \eqref{AFDD1}--\eqref{AFDD2} are \eqref{2d1cKP}.

\paragraph{Darboux transformation}  
Similarly from \eqref{DTFL1}--\eqref{DTFL2}, we get the Darboux transformation for \eqref{2d1cKPLP1}--\eqref{2d1cKPLP2}, either expressed in terms of the $x$-derivative or difference operators $\Delta_i$
\begin{align}
\tilde\phi&=\phi_x-\theta_x\theta^{-1}\phi=\Delta_i\phi-(\Delta_i\theta)\theta^{-1}\phi,\label{DDDTFL1}\\
\tilde G&=G-\theta_x\theta^{-1}=T_iG-(\Delta_i\theta)\theta^{-1},\label{DDDTFL2}
\end{align}
for $i=1,2$.
\paragraph{Binary Darboux transformation}
It is clear that the binary Darboux transformation always keeps the same form as \eqref{HMBDT1}-\eqref{HMBDT3}  whatever the continuum limits are. The only difference is the definition of the eigenfunction potential $\Omega$. Here and hereafter, we will only give the definition of $\Omega$. The leading order terms in \eqref{HMQ} give the eigenfunction potential
\begin{align}
    \Omega_x&=\psi^*\phi,\label{DDQ1}\\
    \Delta_i\Omega&=T_i(\psi^*)\phi,\,\, i=1,2,\label{DDQ2}
\end{align}
\paragraph{Solutions}
From the continuum limit of the quasicasoratian solutions \eqref{DT G} for $i=3$, we get an alternative quasiwronskian expression and so the Darboux transformations give solutions
\begin{equation}\label{DT G2}
    G[n]=G+
    \begin{vmatrix}
        \Theta&0\\
        \vdots&\vdots\\
        \Theta^{(n-2)}&0\\
        \Theta^{(n-1)}&1\\
        \Theta^{(n)}&\fbox{$0$}
    \end{vmatrix}
    =T^n_iG+
        \begin{vmatrix}
            \Theta&0\\
            \vdots&\vdots\\
            \Delta_i^{n-2}\Theta&0\\
            \Delta_i^{n-1}\Theta&1\\
            \Delta_i^n\Theta&\fbox{$0$}
        \end{vmatrix},
\end{equation}
for $i=1,2$. The quasigrammian solutions are given by \eqref{QWGHM} with $\Omega$ defined by \eqref{DDQ1}--\eqref{DDQ2}.

\subsection{Noncommutative $\Delta\partial^2$KP equation}
To take the second continuum limit, let $a_2\to 0$ and note that,
for example, $\Delta_2G=G_x+{1\over 2}a_2(G_{xx}+G_y)+O(a_2^2)$ and
$\nabla_2G=G_x+{1\over 2}a_2(G_y-G_{xx})+O(a_2^2)$.
\paragraph{Nonlinear equation}
The leading order terms in \eqref{2d1cKP} give the noncommutative differential-difference version of the potential KP equation
\begin{equation}
a_1^2\Delta_1G_x\Delta_1G-a_1[\Delta_1G,G_x]=-\Delta_1G_x+\tfrac12a_1(\Delta_1G_{xx}+\Delta_1G_y)+G_{xx}.\label{DDKP}
\end{equation}
\paragraph{Lax pair} The leading order terms in \eqref{2d1cKPLP1}
and \eqref{2d1cKPLP2} give the Lax pair
\begin{eqnarray}
    &&\phi_y+\phi_{xx}=2G_x\phi,\label{DDDLP1}\\
    &&\Delta_1\phi=\phi_x+a_1(\Delta_1 G)\phi.\label{DDDLP2}
\end{eqnarray}
The leading order terms in \eqref{AFDD1}--\eqref{AFDD2} give the adjoint Lax pair
\begin{eqnarray}
    &&\psi_y-\psi_{xx}+2G_x^*\psi=0,\label{ADDLP1}\\
    &&\nabla_1\psi=\psi_x-a_1(\nabla_1 G^*)\psi.\label{ADDLP2}
\end{eqnarray}
The compatibility conditions of both \eqref{DDDLP1}--\eqref{DDDLP2}
and \eqref{ADDLP1}--\eqref{ADDLP2} are \eqref{DDKP}.
\paragraph{Darboux transformation}
Similarly, from \eqref{DDDTFL1}--\eqref{DDDTFL2}, we get the Darboux
transformation for \eqref{DDDLP1}--\eqref{DDDLP2}, either expressed
in terms of $x$-derivative or the difference operator $\Delta_1$
\begin{align}
\tilde{\phi}&=\phi_x-\theta_x\theta^{-1}\phi=\Delta_1\phi-(\Delta_1\theta)\theta^{-1}\phi,\label{DDDT1}\\
\tilde{G}&=G-\theta_x\theta^{-1}.\label{DDDT2}
\end{align}

\paragraph{Binary Darboux transformation}
Considering the expansion of \eqref{DDQ2} for $i=2$ as $a_2\to0$
gives
\[
    \Omega_x+\tfrac12a_2(\Omega_{xx}+\Omega_y)=(\psi^*+a_2\psi^*_x)\phi+O(a_2^2).
\]
From this it follows that $\Omega_y=2\psi_x^*\phi-\Omega_{xx}$ and so
the eigenfunction potential $\Omega$ satisfies
\begin{align}
\Omega_x&=\psi^*\phi,\label{DDDQ1}\\
\Omega_y&=\psi_x^*\theta-\psi^*\theta_x,\label{DDDQ2}\\
\Delta_1\Omega&=(T_1\psi)^*\phi.\label{DDDQ3}
\end{align}

\paragraph{Solutions} The Darboux transformations give solutions
\begin{equation}\label{DT G3}
    G[n]=G+
    \begin{vmatrix}
        \Theta&0\\
        \vdots&\vdots\\
        \Theta^{(n-2)}&0\\
        \Theta^{(n-1)}&1\\
        \Theta^{(n)}&\fbox{$0$}
    \end{vmatrix}
    =T^n_1G+
        \begin{vmatrix}
            \Theta&0\\
            \vdots&\vdots\\
            \Delta_1^{n-2}\Theta&0\\
            \Delta_1^{n-1}\Theta&1\\
            \Delta_1^n\Theta&\fbox{$0$}
        \end{vmatrix},
\end{equation}
and the quasigrammian solutions are given by \eqref{QWGHM} with $\Omega$ defined by \eqref{DDDQ1}--\eqref{DDDQ3}.

\subsection{Noncommutative KP equation}
To take the final continuum limit, let $a_1\to 0$ and note that here
$\Delta_1G=G_x+{1\over 2}a_1(G_y+G_{xx})+{1\over
6}a_1^2(G_{xxx}+3G_{xy}+2G_t)+O(a_1^3)$ and $\nabla_1G=G_x+{1\over
2}a_1(G_y-G_{xx})+\tfrac 16a_1^2(2G_t-3G_{xy}+G_{xxx})+O(a_1^3)$.
\paragraph{Nonlinear equation} The leading order terms in
\eqref{DDKP} give the noncommutative potential KP equation
\begin{equation}
(-4G_t-6G_x^2+G_{xxx})_x+3G_{yy}=6[G_x,G_y].\label{ncKP}
\end{equation}
\paragraph{Lax pair}
The leading order terms in \eqref{DDDLP1}--\eqref{DDDLP2} give the
Lax pair
\begin{align}
    \phi_y&=-\phi_{xx}+2G_x\phi,\label{KPLP1}\\
    \phi_t&=\phi_{xxx}-3G_x\phi_x+\tfrac32(G_y-G_{xx})\phi.\label{KPLP2}
\end{align}
Similarly, the leading order terms in \eqref{ADDLP1}-\eqref{ADDLP2}
give the adjoint form
\begin{align}
    \psi_y&=\psi_{xx}-2G_x^*\psi,\label{cKPLP1}\\
    \psi_t&=\psi_{xxx}-3G_x^*\psi_x-\tfrac32(G_y^*+G_{xx}^*)\psi.\label{cKPLP2}
\end{align}
The compatibility conditions of both \eqref{KPLP1}--\eqref{KPLP2}
and \eqref{cKPLP1}--\eqref{cKPLP2} are \eqref{ncKP}.
\paragraph{Darboux transformation} The leading orders in
\eqref{DDDT1}--\eqref{DDDT2} give the Darboux transformation for
\eqref{ncKP}
\begin{align}
\tilde{\phi}&=\phi_x-\theta_x\theta^{-1}\phi,\\
\tilde{G}&=G-\theta_x\theta^{-1}.
\end{align}
\paragraph{Binary Darboux transformation}
Considering the expansion of \eqref{DDQ2} for $i=2$ as $a_1\to0$
gives
\[
    \Omega_x+\tfrac12a_1(\Omega_{xx}+\Omega_y)+\tfrac 16a_1^2(\Omega_{xxx}+3\Omega_{xy}+2\Omega_t)
    =(\psi^*+a_1\psi^*_x+\tfrac 12a_1^2(\psi_{xx}^*+\psi_y^*))\phi+O(a_2^3).
\]
From this it follows that
$2\Omega_t=3(\psi_{xx}^*+\psi_y^*)\phi-\Omega_{xxx}-3\Omega_{xy}$
and so the eigenfunction potential $\Omega$ satisfies
\begin{align}
\Omega_x&=\psi^*\phi,\label{KPQ1}\\
\Omega_y&=\psi_x^*\phi-\psi^*\phi_x,\label{KPQ2}\\
\Omega_t&=\psi_{xx}^*\phi-\psi_x^*\phi_x+\psi^*\phi_{xx}-3\psi^*G_x\phi.\label{KPQ3}
\end{align}
\begin{rem}
In \cite{GN1}, the following noncommutative KP equation is
considered
\begin{equation}
(U_t+3U_x^2+U_{xxx})_x+3U_{yy}=3[U_x,U_y].\label{ncKP1}
\end{equation}
Its quasideterminant solutions are obtained through Darboux
transformation and binary Darboux transformation. In fact, using the
scaling transformations $y\to-y,\,t\to -4t$ and writing in terms of
$U=-2G$, \eqref{ncKP} can be transformed into \eqref{ncKP1}.

\paragraph{Solutions} Finally, we recover the quasideterminant solutions of the noncommutative KP equation, either quasiwronskian \cite{EGR, GN1} or quasigrammian \cite{GN1}
\[
U[n]=U-2\begin{vmatrix}
    \Theta&0\\
    \vdots&\vdots\\
    \Theta^{(n-2)}&0\\
    \Theta^{(n-1)}&1\\
    \Theta^{(n)}&\fbox{$0$}
\end{vmatrix},\text{ or }U[n]=U-2\begin{vmatrix} \Omega(\Theta, \Rho)&\Rho^*\\
    \Theta&\fbox{$0$}
    \end{vmatrix}.
\]
\end{rem}

\section{Conclusions and discussion}
In this paper, we first obtained the quasigrammian solutions of the non-Abelian Hirota-Miwa equation
by using an iterated binary Darboux transformation. The binary Darboux transformation leads to an ansatz \eqref{ansatz} for $U_{ij}$ which reduces the non-Abelian Hirota-Miwa equations to a single equation symmetric in the three discrete variables $n_i$. Then by considering the continuum limits of this equation and its Darboux transformations and binary Darboux transformations, we obtain a cascade of noncommutative differential-difference equations and their quasideterminant solutions.

There is a second parametrization, $U_{ij}=(a_i^{-1}-a_j^{-1})T_j(C_{i})C_i^{-1}$ which can be seen from the quasiwronskian solution constructed by its Darboux transformation. A natural question is `Can we obtain continuum limits of the non-Abelian Hirota-Miwa equation under the second parametrization?'

Using this parametrization the linear equations are
\begin{equation*}
\Delta_i\phi-\Delta_j\phi=a_j(a_i^{-1}-a_j^{-1})\Delta_j(C_i)C_i^{-1}\phi,\quad
i,j\in\{1,2,3\},
\end{equation*}
and in the limit $a_3\to0$, this system becomes
\begin{eqnarray*}
&&\Delta_1\phi-\Delta_2\phi=a_2(a_1^{-1}-a_2^{-1})\Delta_2(C_1)C_1^{-1}\phi,\\
  &&\Delta_1\phi=\phi_x-C_{1x}C_1^{-1}\phi,\\
    &&\Delta_2\phi=\phi_x-C_{2x}C_2^{-1}\phi.
\end{eqnarray*}
The algebraic compatibility of these equations gives
\begin{equation}
    \label{DDCW}
a_2(a_1-a_2)\Delta_2(C_1)C_1^{-1}+a_1a_2(C_{2x}C_2^{-1}-C_{1x}C_1^{-1})=0,
\end{equation}
and it may be shown also that the only other compatibility condition, $\phi_{,12}=\phi_{,21}$ is identically satisfied provided \eqref{DDCW} is satisfied.  This, as expected, agrees with the leading order terms in the continuum limit of $U_{12}+U_{23}-U_{13}=0$.

Now let $u:=C_{1x}C_1^{-1}$ and $v:=C_{2x}C_2^{-1}$, then \eqref{DDCW} can be rewritten as
\begin{eqnarray}
&&(u-v)_x+(a_1^{-1}-a_2^{-1})a_2\Delta_2(u)
+a_2\Delta_2(u)v-a_1\Delta_1(v)u+[u,v]=0.\label{DDCW16}
\end{eqnarray}
On the other hand, exactly the same equation arises if we write $u=-a_1\Delta_1(G)$ and
$v=-a_2\Delta_2(G)$ in the noncommutative $\Delta^2\partial$KP
equation \eqref{2d1cKP}. Thus we see that in the first continuum limit the second parametrization gives a system equivalent to that obtained from the first. However, it is not at all clear how to take the second continuum limit, $a_2\to0$. For this reason we have not pursued the second parametrization any further.

\section*{Acknowledgement}
This work was supported in part by a Royal Society China Fellowship and
the National Natural Science Foundation of China (grant no.
10601028). K.M.~Tamizhmani wishes to acknowledge the  
NBHM for the project and J.J.C.~Nimmo wishes to acknowledge the support of a Royal
Society of London International Outgoing Short Visit award.


\begin{thebibliography}{99}
\bibitem{K}B.A. Kupershmidt, KP or mKP: Noncommutative Mathematics
of Lagrangian, Hamiltonian, and Integrable systems. Mathematical
Surveys and Monographs (American Mathematical Society, New York),
78, 2000.
\bibitem{P}L.D. Paniak, Exact noncommutative KP and KdV multi-solitons, arXiv:hep-th/0105185
(2001).
\bibitem{S}M. Sakakibara, J. Phys. A 37 (2004) L599-L604.
\bibitem{WW1}N. Wang, M. Wadati, J. Phys. Soc. Jpn 72 (2003)
1366-1373.
\bibitem{WW2}N. Wang, M. Wadati, J. Phys. Soc. Jpn 72 (2003)
1881-1888.
\bibitem{WW3}N. Wang, M. Wadati, J. Phys. Soc. Jpn 73 (2004)
1689-1698.
\bibitem{H} M. Hamanaka, Noncommutative solitons and D-branes. PhD
thesis. arXiv:hep-th/0303256 (2003).
\bibitem{HT1}M. Hamanaka, Nucl. Phys. B 316 (2003) 77-83.
\bibitem{DH}A. Dimakis, M\"uller-Hoissen, J. Phys. A 38 (2005)
5453-5505.
\bibitem{JN}J.J.C. Nimmo, J. Phys. A 39 (2006) 5053-5065.
\bibitem{GN1}C.R. Gilson, J.J.C. Nimmo, J. Phys. A 40 (2007)
3839-3850.
\bibitem{GN2}C.R. Gilson, J.J.C. Nimmo, Y. Ohta, J. Phys. A: Math.
Theor. 40 (2007) 12607-12617.
\bibitem{GNS}C.R. Gilson, J.J.C. Nimmo, C.M. Sooman, J. Phys. A:
Math. Theor. 41 (2008) 085202.
\bibitem{LN}C.X. Li, J.J.C. Nimmo, Proc. R. Soc. A 464 (2008)
951-966.
\bibitem{GR}I.M. Gelfand, V.S. Retakh, Funkt. Anal. Prilozhen. 25
(1991) 13-25.
\bibitem{EGR}P. Etingof, I.M. Gelfand, V.S. Retakh, Math. Res.
Lett. 5(1998) 1-12.
\bibitem{GGRL}I.M. Gelfand, S. Gelfand, V.M. Retakh, R.L. Wilson,
Adv. Math. 193 (2005) 56-141.
\bibitem{HR}R. Hirota, J. Phys. Soc. Japan 50 (1982) 3785-91.
\bibitem{MT}T. Miwa, Proc. Japan Acad. A 58 (1982) 9-12.
\end{thebibliography}
\end{document}